\documentclass[%
 reprint,
 superscriptaddress,
showpacs,preprintnumbers,
 amsmath,
amssymb,
 aps,
floatfix,
]{revtex4-1}

\usepackage{graphicx}
\usepackage{dcolumn}
\usepackage{bm}
\usepackage{color}
\usepackage{xspace}
\usepackage{lineno}
\usepackage{hyperref}
\usepackage{natbib}
\usepackage{ulem}
\input symbols
\linenumbersep{0.1cm}

\begin{document}


\preprint{arXiv:1308.0465v2 [hep-ex] Accepted for Publication by Physical Review Letters}
\title{Measurement of Neutrino Oscillation Parameters from Muon Neutrino Disappearance with an Off-axis Beam}



\newcommand{\INSTC}{\affiliation{University of Alberta, Centre for Particle Physics, Department of Physics, Edmonton, Alberta, Canada}}
\newcommand{\INSTEE}{\affiliation{University of Bern, Albert Einstein Center for Fundamental Physics, Laboratory for High Energy Physics (LHEP), Bern, Switzerland}}
\newcommand{\INSTFE}{\affiliation{Boston University, Department of Physics, Boston, Massachusetts, U.S.A.}}
\newcommand{\INSTD}{\affiliation{University of British Columbia, Department of Physics and Astronomy, Vancouver, British Columbia, Canada}}
\newcommand{\INSTGA}{\affiliation{University of California, Irvine, Department of Physics and Astronomy, Irvine, California, U.S.A.}}
\newcommand{\INSTI}{\affiliation{IRFU, CEA Saclay, Gif-sur-Yvette, France}}
\newcommand{\INSTCI}{\affiliation{Chonnam National University, Institute for Universe \& Elementary Particles, Gwangju, Korea}}
\newcommand{\INSTGB}{\affiliation{University of Colorado at Boulder, Department of Physics, Boulder, Colorado, U.S.A.}}
\newcommand{\INSTFG}{\affiliation{Colorado State University, Department of Physics, Fort Collins, Colorado, U.S.A.}}
\newcommand{\INSTCJ}{\affiliation{Dongshin University, Department of Physics, Naju, Korea}}
\newcommand{\INSTFH}{\affiliation{Duke University, Department of Physics, Durham, North Carolina, U.S.A.}}
\newcommand{\INSTBA}{\affiliation{Ecole Polytechnique, IN2P3-CNRS, Laboratoire Leprince-Ringuet, Palaiseau, France }}
\newcommand{\INSTEF}{\affiliation{ETH Zurich, Institute for Particle Physics, Zurich, Switzerland}}
\newcommand{\INSTEG}{\affiliation{University of Geneva, Section de Physique, DPNC, Geneva, Switzerland}}
\newcommand{\INSTDG}{\affiliation{H. Niewodniczanski Institute of Nuclear Physics PAN, Cracow, Poland}}
\newcommand{\INSTCB}{\affiliation{High Energy Accelerator Research Organization (KEK), Tsukuba, Ibaraki, Japan}}
\newcommand{\INSTED}{\affiliation{Institut de Fisica d'Altes Energies (IFAE), Bellaterra (Barcelona), Spain}}
\newcommand{\INSTEC}{\affiliation{IFIC (CSIC \& University of Valencia), Valencia, Spain}}
\newcommand{\INSTEI}{\affiliation{Imperial College London, Department of Physics, London, United Kingdom}}
\newcommand{\INSTGF}{\affiliation{INFN Sezione di Bari and Universit\`a e Politecnico di Bari, Dipartimento Interuniversitario di Fisica, Bari, Italy}}
\newcommand{\INSTBE}{\affiliation{INFN Sezione di Napoli and Universit\`a di Napoli, Dipartimento di Fisica, Napoli, Italy}}
\newcommand{\INSTBF}{\affiliation{INFN Sezione di Padova and Universit\`a di Padova, Dipartimento di Fisica, Padova, Italy}}
\newcommand{\INSTBD}{\affiliation{INFN Sezione di Roma and Universit\`a di Roma ``La Sapienza'', Roma, Italy}}
\newcommand{\INSTEB}{\affiliation{Institute for Nuclear Research of the Russian Academy of Sciences, Moscow, Russia}}
\newcommand{\INSTHA}{\affiliation{Kavli Institute for the Physics and Mathematics of the Universe (WPI), Todai Institutes for Advanced Study, University of Tokyo, Kashiwa, Chiba, Japan}}
\newcommand{\INSTCC}{\affiliation{Kobe University, Kobe, Japan}}
\newcommand{\INSTCD}{\affiliation{Kyoto University, Department of Physics, Kyoto, Japan}}
\newcommand{\INSTEJ}{\affiliation{Lancaster University, Physics Department, Lancaster, United Kingdom}}
\newcommand{\INSTFC}{\affiliation{University of Liverpool, Department of Physics, Liverpool, United Kingdom}}
\newcommand{\INSTFI}{\affiliation{Louisiana State University, Department of Physics and Astronomy, Baton Rouge, Louisiana, U.S.A.}}
\newcommand{\INSTJ}{\affiliation{Universit\'e de Lyon, Universit\'e Claude Bernard Lyon 1, IPN Lyon (IN2P3), Villeurbanne, France}}
\newcommand{\INSTCE}{\affiliation{Miyagi University of Education, Department of Physics, Sendai, Japan}}
\newcommand{\INSTDF}{\affiliation{National Centre for Nuclear Research, Warsaw, Poland}}
\newcommand{\INSTFJ}{\affiliation{State University of New York at Stony Brook, Department of Physics and Astronomy, Stony Brook, New York, U.S.A.}}
\newcommand{\INSTGJ}{\affiliation{Okayama University, Department of Physics, Okayama, Japan}}
\newcommand{\INSTCF}{\affiliation{Osaka City University, Department of Physics, Osaka, Japan}}
\newcommand{\INSTGG}{\affiliation{Oxford University, Department of Physics, Oxford, United Kingdom}}
\newcommand{\INSTBB}{\affiliation{UPMC, Universit\'e Paris Diderot, CNRS/IN2P3, Laboratoire de Physique Nucl\'eaire et de Hautes Energies (LPNHE), Paris, France}}
\newcommand{\INSTGC}{\affiliation{University of Pittsburgh, Department of Physics and Astronomy, Pittsburgh, Pennsylvania, U.S.A.}}
\newcommand{\INSTFA}{\affiliation{Queen Mary University of London, School of Physics and Astronomy, London, United Kingdom}}
\newcommand{\INSTE}{\affiliation{University of Regina, Department of Physics, Regina, Saskatchewan, Canada}}
\newcommand{\INSTGD}{\affiliation{University of Rochester, Department of Physics and Astronomy, Rochester, New York, U.S.A.}}
\newcommand{\INSTBC}{\affiliation{RWTH Aachen University, III. Physikalisches Institut, Aachen, Germany}}
\newcommand{\INSTDD}{\affiliation{Seoul National University, Department of Physics and Astronomy, Seoul, Korea}}
\newcommand{\INSTFB}{\affiliation{University of Sheffield, Department of Physics and Astronomy, Sheffield, United Kingdom}}
\newcommand{\INSTDI}{\affiliation{University of Silesia, Institute of Physics, Katowice, Poland}}
\newcommand{\INSTEH}{\affiliation{STFC, Rutherford Appleton Laboratory, Harwell Oxford,  and  Daresbury Laboratory, Warrington, United Kingdom}}
\newcommand{\INSTCH}{\affiliation{University of Tokyo, Department of Physics, Tokyo, Japan}}
\newcommand{\INSTBJ}{\affiliation{University of Tokyo, Institute for Cosmic Ray Research, Kamioka Observatory, Kamioka, Japan}}
\newcommand{\INSTCG}{\affiliation{University of Tokyo, Institute for Cosmic Ray Research, Research Center for Cosmic Neutrinos, Kashiwa, Japan}}
\newcommand{\INSTGI}{\affiliation{Tokyo Metropolitan University, Department of Physics, Tokyo, Japan}}
\newcommand{\INSTF}{\affiliation{University of Toronto, Department of Physics, Toronto, Ontario, Canada}}
\newcommand{\INSTB}{\affiliation{TRIUMF, Vancouver, British Columbia, Canada}}
\newcommand{\INSTG}{\affiliation{University of Victoria, Department of Physics and Astronomy, Victoria, British Columbia, Canada}}
\newcommand{\INSTDJ}{\affiliation{University of Warsaw, Faculty of Physics, Warsaw, Poland}}
\newcommand{\INSTDH}{\affiliation{Warsaw University of Technology, Institute of Radioelectronics, Warsaw, Poland}}
\newcommand{\INSTFD}{\affiliation{University of Warwick, Department of Physics, Coventry, United Kingdom}}
\newcommand{\INSTGE}{\affiliation{University of Washington, Department of Physics, Seattle, Washington, U.S.A.}}
\newcommand{\INSTGH}{\affiliation{University of Winnipeg, Department of Physics, Winnipeg, Manitoba, Canada}}
\newcommand{\INSTEA}{\affiliation{Wroclaw University, Faculty of Physics and Astronomy, Wroclaw, Poland}}
\newcommand{\INSTH}{\affiliation{York University, Department of Physics and Astronomy, Toronto, Ontario, Canada}}

\INSTC
\INSTEE
\INSTFE
\INSTD
\INSTGA
\INSTI
\INSTCI
\INSTGB
\INSTFG
\INSTCJ
\INSTFH
\INSTBA
\INSTEF
\INSTEG
\INSTDG
\INSTCB
\INSTED
\INSTEC
\INSTEI
\INSTGF
\INSTBE
\INSTBF
\INSTBD
\INSTEB
\INSTHA
\INSTCC
\INSTCD
\INSTEJ
\INSTFC
\INSTFI
\INSTJ
\INSTCE
\INSTDF
\INSTFJ
\INSTGJ
\INSTCF
\INSTGG
\INSTBB
\INSTGC
\INSTFA
\INSTE
\INSTGD
\INSTBC
\INSTDD
\INSTFB
\INSTDI
\INSTEH
\INSTCH
\INSTBJ
\INSTCG
\INSTGI
\INSTF
\INSTB
\INSTG
\INSTDJ
\INSTDH
\INSTFD
\INSTGE
\INSTGH
\INSTEA
\INSTH

\author{K.\,Abe}\INSTBJ
\author{J.\,Adam}\INSTFJ
\author{H.\,Aihara}\INSTCH\INSTHA
\author{T.\,Akiri}\INSTFH
\author{C.\,Andreopoulos}\INSTEH
\author{S.\,Aoki}\INSTCC
\author{A.\,Ariga}\INSTEE
\author{T.\,Ariga}\INSTEE
\author{S.\,Assylbekov}\INSTFG
\author{D.\,Autiero}\INSTJ
% ...
"\author{Y.\,Kudenko}\thanks{affiliated member at Moscow Institute of Physics and Technology and National
Research Nuclear University, Russia}\INSTEB
\author{S.\,Kumaratunga}\INSTB
\author{R.\,Kurjata}\INSTDH
\author{T.\,Kutter}\INSTFI
\author{J.\,Lagoda}\INSTDF
\author{K.\,Laihem}\INSTBC
\author{M.\,Laveder}\INSTBF
\author{M.\,Lawe}\INSTFB
\author{M.\,Lazos}\INSTFC
\author{K.P.\,Lee}\INSTCG
% ...

\collaboration{The T2K Collaboration}\noaffiliation


\date{\today}

\begin{abstract}


The T2K collaboration reports a precision measurement of muon neutrino disappearance with an off-axis neutrino beam with a peak energy of 0.6 GeV.  Near detector measurements are used to constrain the neutrino flux and cross section parameters. The Super-Kamiokande far detector, which is 295 km downstream of the neutrino production target, collected data corresponding to $3.01 \times 10^{20}$ protons on target.
In the absence of neutrino oscillations,
$205 \pm 17$ (syst.) events are expected to be detected while only 58 muon neutrino event candidates are observed. 
A fit to the neutrino rate and energy spectrum assuming three neutrino flavors 
and normal mass hierarchy 
yields a best-fit mixing angle
$\sin^2(\theta_{23})  = 0.514\pm0.082$
 and mass splitting 
$|\Delta m^2_{32}| = 2.44^{+0.17}_{-0.15}\times 10^{-3} $ eV$^2$/c$^4$.
Our result corresponds to the maximal oscillation disappearance probability.

\end{abstract}
\maketitle

{\it Introduction.}\textemdash 
Oscillations between different neutrino flavor states are a physics process well described by the $3\times3$ Pontecorvo-Maki-Nakagawa-Sakata mixing 
matrix~\cite{Pont1,*Pont2,*Pont3,*MNS}, 
which is parametrized~\cite{MCMC_ref} by three mixing angles $\theta_{12}$,  $\theta_{23}$, $\theta_{13}$, and a CP violating phase $\delta_{CP}$.   In this mixing scheme, the angle $\theta_{23}$ and mass splitting 
$\Delta m^2_{32}$ are the main parameters that govern atmospheric and long-baseline $\nu_\mu$ disappearance oscillations. 
The oscillation probability in the limit $|\Delta m^2_{32}|\gg|\Delta m^2_{21}|$ is 
\begin{linenomath}
\begin{align}
P(\nu_\mu \rightarrow \nu_\mu) \simeq& 
    1-4 \cos^2(\theta_{13})\sin^2(\theta_{23}) 
  [1-\cos^2(\theta_{13})\nonumber \\
   &\times \sin^2(\theta_{23})] 
\sin^2(1.27\Delta m^2_{32}L/E_\nu),
  \label{eq:oscprob}
\end{align}
\end{linenomath}
where $L(\mathrm{km})$ is the neutrino propagation distance, $E_\nu(\mathrm{GeV})$ is the neutrino energy
and $\Delta m^2_{32} (\mathrm{eV^2})$ is the neutrino mass splitting. 
Recent 
measurements 
\cite{PhysRevD.85.031103,PhysRevD.81.092004,Adamson:2012rm,minos-apr2013} 
are consistent with maximal $\nu_\mu$ disappearance
for which $\theta_{23}$ is approximately $\pi /4$.
Improved knowledge of this angle has an important impact on neutrino mass models and on the interpretation of the $\nu_e$ appearance results, given the recent findings of non-zero
$\theta_{13}$ 
measurements~\cite{PhysRevLett.107.041801, *Adamson:2011qu,  *PhysRevLett.108.171803,
*PhysRevLett.108.131801, 
*PhysRevLett.108.191802}.
In this paper, we report on new measurements on the values of $\sin^2(\theta_{23})$
and $|\Delta m^2_{32}|$.

{\it T2K Experiment.}\textemdash 
The T2K experiment~\cite{Abe:2011ks} uses a 30 GeV proton
beam from the J-PARC accelerator facility. This
combines (1) a muon neutrino beam line, 
(2) the near detector complex, which is located 280 m downstream of the neutrino production target and measures the neutrino beam, which constrains the neutrino flux parametrization and cross sections, and (3) the far detector, Super-Kamiokande (SK), which detects neutrinos at a baseline distance of $L=295$ km from the target. The neutrino beam 
is directed 2.5$^\circ$ away from SK producing a narrow-band $\nu_{\mu}$ beam~\cite{PhysRevD.87.012001} at the far detector
whose energy peaks at  
$E_\nu$=$\Delta m^2_{32}L/2\pi$ $\approx 0.6$~GeV 
which corresponds to the first oscillation minimum of the $\nu_\mu$ survival probability at SK. 
This enhances the sensitivity to determine $\theta_{23}$ from the oscillation
measurements and reduces backgrounds from higher-energy neutrino interactions at SK.

The J-PARC main ring accelerator produces a fast-extracted  proton beam. 
 The primary beam line has 21 electrostatic beam position monitors, 19 secondary emission monitors, an optical transition radiation monitor, and five current transformers which measure the proton current before a graphite target.
Pions and kaons produced in the target decay in the secondary beam line, 
which contains three focusing horns and a 96-m-long decay tunnel. This is followed by a beam dump and
a set of muon monitors.

The near detector complex contains an on-axis Interactive Neutrino Grid detector (INGRID)~\cite{Abe2012} and an off-axis magnetic detector, ND280.
A schematic detector layout is published elsewhere~\cite{Abe:2011ks}.  
The INGRID detector has 14 seven-ton iron-scintillator tracker modules arranged in a 10-m horizontal by 10-m vertical crossed array. This detector provides high-statistics monitoring of the beam intensity, direction, profile, and stability.  The off-axis detector is enclosed in a 0.2-T magnet that contains 
a subdetector optimized to measure $\pi^0$s (P$\O$D)~\cite{Assylbekov201248},
three time projection chambers (TPC1,2,3)~\cite{Abgrall:2010hi} 
alternating with two one-ton fine grained detectors (FGD1,2)~\cite{Amaudruz:2012pe}, 
and an electromagnetic calorimeter (ECal)~\cite{allan2013electromagnetic} that surrounds the TPC, FGD, and P$\O$D detectors. A side muon range detector (SMRD)~\cite{Aoki:2012mf}, built into slots in the magnet flux return steel, identifies muons that exit or 
stop in the magnet steel when the path length exceeds the energy loss range. 

The SK water Cherenkov far detector~\cite{Ashie:2005ik} has a 22.5 kt fiducial volume within a cylindrical inner detector (ID) instrumented with 11129 inward facing 20-inch phototubes. Surrounding the ID is a 2-meter wide outer detector (OD) with 1885 outward-facing 8-inch phototubes. A Global Positioning System with $<$150 ns precision synchronizes the timing between SK events and the J-PARC  beam spill.   
 
These results are based on three periods: Run 1 (January-June 2010), Run 2 (November 2010-March 2011), and Run 3 (January-June 2012). 
The proton beam power on the target steadily increased from Run 1, reaching 200 kW with about $10^{14}$ protons per pulse on the target by the end of Run 3.
The total neutrino beam exposure on the SK detector 
corresponds to an integrated $3.01 \times 10^{20}$ protons on target (POT). 

{\it Analysis Strategy.}\textemdash 
The analysis method  
estimates oscillation parameters by comparing the observed and
predicted $\nu_\mu$ interaction rate and energy spectrum at the far detector. 
The rate and spectrum
depend on the oscillation parameters, the incident neutrino flux, 
neutrino interaction cross sections,
and the detector response. 
The initial estimate of the neutrino flux is determined by detailed simulations incorporating
proton beam measurements, INGRID measurements, 
and the pion and kaon production measured by the NA61/SHINE~\cite{Abgrall:2011ae, *Abgrall:2011ts} 
experiment. The ND280 detector measurement of  $\nu_\mu$ charged current (CC) events constrains the initial flux estimates and 
parameters of the neutrino interaction models that affect the predicted rate and 
spectrum of neutrino interactions at both ND280 and SK. 
At SK,  $\nu_\mu$ charged current quasi-elastic (CCQE) events are selected and efficiencies are determined, along with their dependence on final state interactions (FSI) inside the nucleus 
and secondary pion interactions (SI) in the detector material.
These are used in a binned likelihood ratio fit 
to determine the oscillation parameters.  


{\it Initial Neutrino Flux Model.}\textemdash 
To predict the neutrino flux at the near and far detectors, 
the interactions of the primary beam protons and subsequent secondary particles in a graphite target
are modeled with a FLUKA2008~\cite{Ferrari:2005zk, *Battistoni:2007zzb} 
simulation. 
GEANT3~\cite{GEANT3}  
simulations model the secondary particles  
in the magnetic horns and the decay region, and their decays into
neutrinos. 
The hadron interactions  are modeled with 
GCALOR~\cite{GCALOR}. 
The simulation is tuned using measurements of the primary proton beam profile and 
the T2K horn magnetic fields and the NA61/SHINE hadron production
results~\cite{Abgrall:2011ae, *Abgrall:2011ts}.  
The beam direction and neutrino rate per proton on target
are monitored continuously with INGRID, and 
the variations are 
less than
the assigned systematic uncertainties~\cite{nue2013}.
The 
uncertainties 
in 
the flux are 10-20\% in the relevant energy range, dominated by the hadron production uncertainties.
The detailed flux calculations are described elsewhere~\cite{PhysRevD.87.012001}. 

{\it Neutrino Interaction Simulations and 
Cross Section Parameters}\textemdash
 Neutrino interactions in the ND280 and SK detectors are simulated with the NEUT Monte Carlo generator
\cite{Hayato:2009}. External data, primarily from the MiniBooNE experiment~\cite{mb-ccqe}, are used to tune some NEUT neutrino interaction parameters. These determine the input parameter uncertainties used in the fit to the ND280 data~\cite{nue2013}.
Neutrino interaction parameters fall into two categories: parameters 
that are common
between ND280 and SK, and independent parameters affecting interactions at only one of the detectors. 
The common parameters include the axial masses for CCQE and resonant pion production, as well as 5 energy dependent normalizations; these are included in the fit to the ND280 data, which is discussed in the next section.
Since the ND280 target is mainly carbon and differs from the SK target which is mainly oxygen, 
additional independent parameters are required. 
These affect the nuclear model for CCQE (Fermi momentum, 
binding energy and spectral function modeling) and include 
five cross section parameters related to pion production, the neutral current (NC) cross section, the $\nu_e / \nu_\mu$ CC cross section ratio, and the $\nu / \bar{\nu}$ CC cross section ratio.  These 
independent 
cross section uncertainties (11 parameters) produce a 6.3\% fractional error in the expected number of SK events as listed in Table~\ref{tab:nsk_systematic_table_summary}.
Not simulated by NEUT are multi-nucleon knock-out 
processes~\cite{Nieves,*Martini:2010}
that may affect 
~\cite{Meloni,*Lalakulich3,*Martini2,*Martini3} 
oscillation parameter determination strongly.
Our estimation of the bias on the oscillation parameters from 
these processes 
appears to be smaller than the current statistical precision. 

{\it ND280 Measurements, Flux and 
Common 
Cross Section fits.}\textemdash The ND280 detector measures 
inclusive
CC events with a vertex in  FGD1 located upstream of FGD2 and with the 
muon passing through TPC2. The event selection uses the highest-momentum negatively
charged track entering
TPC2 that matches a vertex inside the upstream FGD1 fiducial volume. In addition, the
measured track energy loss in TPC2 must be compatible with a muon.
Events originating from interactions in upstream detectors are
vetoed by excluding events with a track in the TPC1 upstream of FGD1.
This suppresses events with interactions occuring upstream of FGD1
or with a charged particle going backwards from FGD1 into TPC1. 
Using an inclusive CC selection, the efficiency is 47.6\% with a purity
of 88.1\%. 
The main backgrounds are events where the neutrino interactions occur outside
FGD1 and migrate into the fiducial volume due to mis-reconstruction,
or from neutral particles interacting within the FGD1.

The CC inclusive sample is further subdivided into two samples
called CCQE and CCnQE. The CCQE sample is optimized to select charged current quasi-elastic
events and the CCnQE sample contains the remaining events. 
This separation is made to improve constraints
on the neutrino flux and cross section parameters.  
The CCQE selection vetoes events with additional tracks
that cross FGD1 and TPC2 or have electrons from muon
decay found inside FGD1. After beam and data quality cuts, there
are 5841 CCQE and 5214 CCnQE events 
that correspond to an integrated dataset of
$2.66\times10^{20}$ POT.
These two data selections are each subdivided into 
5(momentum) $\times$ 4(angular) 
bins which produces a 40-bin histogram used in a fit to the ND280 data.

\begin{figure}
\includegraphics[width=3.25in]{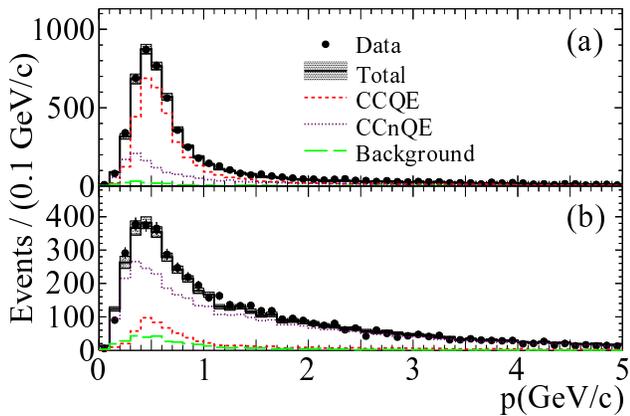}
\caption{\label{fig:nd280pmu-inc} 
The ND280 momentum data distributions of 
(a) the CCQE and (b) CCnQE selections. 
The predicted total, CCQE, CCnQE and background event distributions from the ND280 fit 
are overlaid on both figures.
}
\end{figure}

The 40-bin histogram and cosmic ray control samples are fit to estimate the neutrino flux 
crossing ND280 in 11 bins of $E_{\nu_\mu}$, 7 common and 4 ND280 neutrino interaction parameters, 
detector response parameters, and their covariance. 
This ND280 fit also estimates the SK flux parameters, which are constrained through their 
prior covariance with the ND280 flux parameters as calculated by the beam simulation 
described earlier.
The absolute
track momentum scale, pion secondary interactions, and background uncertainties are the largest detector systematics.
The reconstructed ND280 $\mu^-$ momentum distributions for CCQE and CCnQE selections
and predicted event distributions from the ND280 fit to data are shown in Fig.~\ref{fig:nd280pmu-inc}. 
For the oscillation fits, the ND280 fit provides a systematic parameter error matrix which consists of 
11 $E_{\nu_\mu}$ SK flux normalizations, 5 $E_{\bar{\nu}_\mu}$ SK flux normalizations 
and the 7 common neutrino interaction parameters. 
The fractional error on the predicted number of SK candidate events from the
uncertainties in these 23 parameters, as shown in
Table~\ref{tab:nsk_systematic_table_summary}, is 4.2\%.
Without the constraint from the ND280 measurements this fractional
error would be 21.8\%.

 
{\it SK Measurements.}\textemdash 
The SK far detector $\nu_\mu$ candidate events 
are selected from fully-contained beam events.  The SK phototube hits must be within $\pm$500 $\mu$s of the expected neutrino arrival time, and 
there must be low outer detector activity to reject entering background. 
The events must 
satisfy:  visible energy {$>30$~MeV}, exactly one reconstructed Cherenkov ring, 
$\mu$-like particle ID, reconstructed muon momentum {$>200$~MeV}, and 
$\le 1$ reconstructed decay electron.  The reconstructed vertex must 
be in the fiducial volume (at least 2~m away from the ID walls)
and ``flasher'' (intermittent light-emitting phototube) 
events are rejected.  
More details about the SK event selection and reconstruction are found elsewhere~\cite{Ashie:2005ik}.  

\begin{figure}
  \includegraphics[width=8cm]{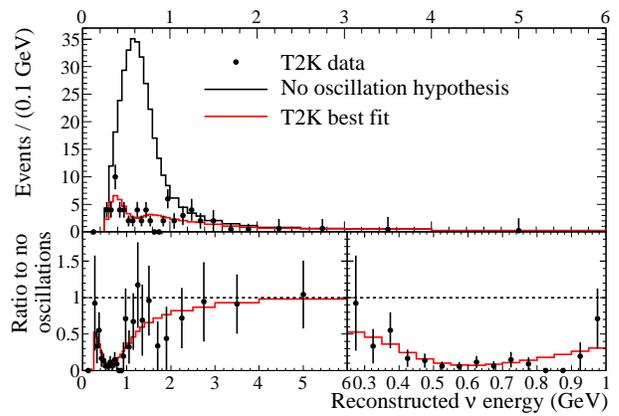} 
  \caption{\label{fig:skspectrum} 
    The 58 event 1-ring $\mu$-like SK reconstructed energy spectrum. 
    Top: The two predicted curves are the no 
    oscillation hypothesis and the best fit from the primary oscillation analysis. 
	The energy scale is given on the top (0-6) GeV.
    Bottom: The ratio of the observed spectrum over the no oscillation hypothesis 
    and ratio of the best fit curve over the no oscillation hypothesis in 
    two energy ranges lower left (0-6) GeV and lower right (0.3-1.0) GeV.
    The fit uses finer binning than is shown here.
  }
\end{figure}

Assuming a quasi-elastic interaction with a bound neutron and neglecting the Fermi motion,  
the neutrino energy is deduced from the detected muon and given by
\begin{linenomath}
\begin{align}
E_{\rm reco} = \frac{ m_p^2-(m_n-E_b)^2 - m_\mu^2+ 2(m_n-E_b)E_\mu}{2(m_n-E_b-E_\mu + p_\mu \cos \theta_\mu)},
\end{align}
\end{linenomath}
where 
 $p_\mu$, $E_\mu$, and $\theta_\mu$ are the reconstructed
muon momentum, energy, and 
the angle with respect to the beam direction, respectively; $m_p$, $m_n$, and $m_\mu$ are masses of the proton, neutron, and muon, respectively, and $E_b=27$~MeV 
is the average binding energy of a nucleon in $^{16}$O. The $E_{\rm reco}$ distribution of the 58 events satisfying the selection criteria is shown in Fig.~\ref{fig:skspectrum}.  The  no-oscillation hypothesis prediction is the solid line in  Fig.~\ref{fig:skspectrum} and the
MC expectation is  $205 \pm 17$ (syst.) events, of which 77.7\% are
$\nu_\mu$+$\bar{\nu}_\mu$ CCQE, 20.7\% are $\nu_\mu$+$\bar{\nu}_\mu$ CCnQE, 1.6\% are NC and 0.02\% are $\nu_e$+$\bar{\nu}_e$ CC.
The expected resolution on reconstructed energy for $\nu_\mu$+$\bar{\nu}_\mu$ CCQE events around the oscillation maximum is $\sim$0.1~GeV.

Eight SK detector systematic uncertainties are associated with event selection and reconstruction. 
The  SK energy scale uncertainty 
is evaluated by comparing energy loss in data and MC for samples of cosmic-ray stopping muons and associated decay-electrons, as well as by comparing reconstructed invariant mass for data and MC for  $\pi^0$s  produced by atmospheric neutrinos.  
The other seven SK event-selection-related uncertainties are also evaluated by comparing atmospheric neutrino MC 
and data samples.  
The $\nu_\mu$+$\bar{\nu}_\mu$ CCQE
ring-counting-based selection uncertainty is evaluated in 
three energy bins, including correlations between energy bins.
Other uncertainties result from selection criteria on the 
$\nu_\mu$+$\bar{\nu}_\mu$ CCQE,  
$\nu_\mu$+$\bar{\nu}_\mu$ CCnQE,
$\nu_e$+$\bar{\nu}_e$ CC, 
and NC events.
These uncertainties (8 parameters) produce a 10.1\% fractional error on the expected number of SK events,
as listed in Table~\ref{tab:nsk_systematic_table_summary}.  

Systematic uncertainties on pion interactions in the target nucleus (FSI) and SK detector (SI) are
evaluated by varying underlying pion scattering cross sections in the NEUT and SK detector simulations.
These uncertainties are evaluated 
separately for $\nu_\mu$+$\bar{\nu}_\mu$ CCQE in three energy bins, 
$\nu_\mu$+$\bar{\nu}_\mu$ CCnQE, $\nu_e$+$\bar{\nu}_e$ CC, and NC events.
The total FSI+SI  uncertainty (6 parameters) on the predicted SK event rate is 3.5\%
as listed in Table~\ref{tab:nsk_systematic_table_summary}.


\def\stst        {\ensuremath{\textrm{sin}^2(2\theta_{12})}\xspace} 
\def\mdmsq     {\ensuremath{|\Delta m^{2}_{32}|\xspace}}
\def\evsqc    {\ensuremath{\rm \,eV^{2} / c^{4}\xspace}}
\def\sysp        {\ensuremath{\boldsymbol{f}}}
\def\nskexp    {\ensuremath{n_{\textrm{SK}}^{\textrm{exp}}}}
\def\nskobs    {\ensuremath{n_{\textrm{SK}}^{\textrm{obs}}}}
\def\nexp    {\ensuremath{n^{\textrm{exp}}}}
\def\nobs    {\ensuremath{n^{\textrm{obs}}}}
\def\LLnd       {\ensuremath{{\cal L}_{\textrm{ND280}}}} 
\def\LLsk       {\ensuremath{{\cal L}_{\textrm{SK}}}} 
\def\pion  {\ensuremath{\pi}\xspace}
\def\cmv  {\ensuremath{{\rm \,cm}^3}\xspace}


\begin{table}
    \centering
    \begin{tabular}{l | D{.}{.}{-1}}
      \hline \hline
      {\bf Source of uncertainty} (no. of parameters)  & \multicolumn{1}{c}{$\delta \nskexp$ / $\nskexp$} \\
      \hline
      ND280-independent cross section (11)              &  6.3\%  \\
      Flux \& ND280-common cross section (23)     &4.2\%\\
      \sk detector systematics (8)                                   & 10.1\% \\
      Final-state and secondary interactions (6)  &  3.5\% \\
      \hline
      Total (48) & 13.1\%  \\
      \hline \hline
    \end{tabular}
  \caption{\label{tab:nsk_systematic_table_summary}
    Effect of 1$\sigma$ systematic parameter variation on
    the number of 1-ring $\mu$-like events,
    computed
    for oscillations with 
    $\sin^2(\theta_{23})=0.500$ and $\mdmsq = 2.40 \times 10^{-3} \evsqc$.
  }
\end{table}



{\it Oscillation Fits.}\textemdash 
The oscillation parameters are
estimated using a binned likelihood ratio to fit
the SK spectrum in the parameter space of
$\sin^2(\theta_{23})$, \mdmsq, and all 48 systematic parameters, $\sysp$,
by minimizing
\begin{linenomath}
\begin{align}
  \displaystyle
  \chi^{2} &
  (\sin^2(\theta_{23}),\mdmsq; \sysp) = 
  (\sysp-{\bf f_{0}})^{T} \cdot {\bf C}^{-1} \cdot (\sysp-{\bf f_{0}}) \nonumber \\
  &+
  2 \sum_{i=1}^{73} 
  \nobs_{i} \textrm{ln}(\nobs_{i}/\nexp_{i}) +
  (\nexp_{i}-\nobs_{i}).
  \label{eq:chisquare:likeratio}
\end{align}
\end{linenomath}
${\bf f_{0}}$ is a 48-dimensional vector with the prior values of 
the systematics parameters, 
${\bf C}$ is the $48 \times 48$ systematic parameter covariance matrix,
$\nobs_{i}$ is the observed number of events in the $i^{th}$ bin and
$\nexp_{i} = \nexp_{i}(\sin^2(\theta_{23}),\mdmsq; \sysp)$ 
is the corresponding expected number of events. 
The sum is over 73 variable-width energy bins, with finer binning in the oscillation peak region.
Oscillation probabilities are calculated using the full three neutrino
oscillation framework.
Normal mass hierarchy is assumed, 
matter effects are included
with an Earth density of $\rho = 2.6 \gm / \cmv$ \cite{Hagiwara:2011kw},
and other oscillation parameters are fixed at the 2012 PDG recommended values~\cite{MCMC_ref}
($\stot = 0.098,
\dmsqso = 7.5 \times 10^{-5} \evsqc, \stst = 0.857$), and with $\dcp = 0$.

The fit to the 58 events using Eq. \ref{eq:chisquare:likeratio}
yields the 
best-fit point 
at
 $\sin^2(\theta_{23})  = 0.514\pm0.082$ 
and 
$\mdmsq = 2.44^{+0.17}_{-0.15}\times 10^{-3} \evsqc$,
 with $\chi^2/ndf= 56.03/71$.
The best-fit neutrino energy spectrum is shown in Fig. \ref{fig:skspectrum}.
The point estimates of the 48 nuisance parameters are all within 0.35 standard deviations of their prior values. 
This fit result value combined with  $\sin^2(2\theta_{13})= 0.098$
corresponds to the maximal possible oscillation disappearance probability
where $\cos^2(\theta_{13})\sin^2(\theta_{23})=0.5$.



The 2D confidence regions for the oscillation parameters 
$\sin^2(\theta_{23}) $ and $\mdmsq$ are constructed using the
constant $\Delta \chi^2$ method~\cite{MCMC_ref}.
The 68\% and 90\% contour regions are shown in Fig. 3. 
Also shown in this figure are the 1D profile
likelihoods for each oscillation parameter separately. 

\begin{figure}
 {\includegraphics[width=8.5cm]{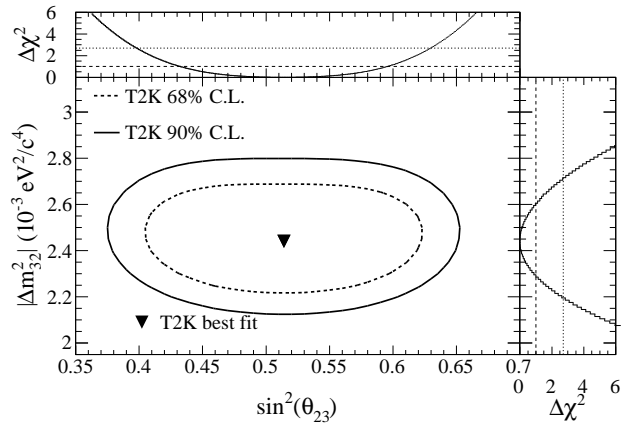} } 
  \caption{\label{fig:contour:analyses} 
    The  68\% and 90\% C.L. contour regions for $\sin^2(\theta_{23})$ and $\mdmsq$ are shown
    for the primary analysis. 
    The 1D profile likelihoods for each oscillation parameter separately
    are also shown. 
}
\end{figure}


An alternative analysis employing a maximum likelihood fit was
performed with the following likelihood function:
\begin{linenomath}
\begin{align}
  \displaystyle
  \LL = &
  \LLnorm(\sin^2(\theta_{23}),\mdmsq,\sysp) \nonumber \\
  &
  \times \LLshape(\sin^2(\theta_{23}),\mdmsq,\sysp) 
  \LLsyst(\sysp),
  \label{eq:chisquare:maxlike}
\end{align}
\end{linenomath}
where $\LLnorm$ is the Poisson probability for the observed number of events,
$\LLshape$ is the likelihood for the reconstructed energy spectrum,
and $\LLsyst$ is analogous to the first term in Eq. \ref{eq:chisquare:likeratio}.
The best-fit point
is at
 $\sin^2(\theta_{23})  = 0.514 $ and $\mdmsq = 2.44\times 10^{-3} \evsqc$.
The primary and alternative analyses are consistent;
the binned maximum fractional difference between best-fit spectra is 1.8\%,
and the confidence regions are almost identical.


A complementary analysis was performed, using Markov Chain Monte Carlo~\cite{MCMC_ref} methods to 
produce a sample of points in the full parameter space distributed according to the posterior probability density.
This analysis
uses both ND280 and SK data simultaneously, rather than separately fitting
the ND280 and SK measurements; the likelihood is the product of the ND280 and SK likelihoods, with the shared systematics treated jointly. 
The  maximum probability density
is found to be $\sin^2(\theta_{23})  = 0.516$ and $\mdmsq = 2.46 \times 10^{-3}
\evsqc$, using a uniform prior probability distribution in both $\sin^2(\theta_{23})$ and $\mdmsq$.
The contours from this analysis are similar in shape and size to the two previously described analyses, but are not expected to be identical due to the difference between Bayesian and classical intervals.
This analysis also has similar results to the ND280 data fit described previously and provides a cross check.


\begin{figure}
 {\includegraphics[width=8.5cm]{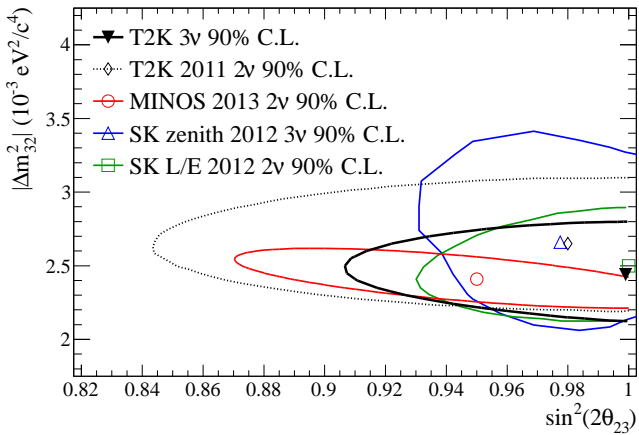} }
  \caption{\label{fig:contour:external} 
    The 90\% C.L. contour region for \sttt and \mdmsq~for the primary T2K analysis
is calculated by the profiling over the octant.
    The T2K 2011\cite{PhysRevD.85.031103},
    SK\cite{Itow201379}, 
    and
    MINOS\cite{minos-apr2013} 90\% C.L. contours
    with different flavor assumptions are shown for comparison. } 
\end{figure}

{\it Conclusions.}\textemdash
The T2K primary result (90\%~C.L. region) is
consistent with maximal mixing and compared 
to other recent experimental results 
in Figure~\ref{fig:contour:external}. 
In this paper the $\nu_\mu$ disappearance analysis, 
based on the $3.01\times 10^{20}$ POT off-axis beam exposure, has
a best-fit mass splitting of 
$\mdmsq = 2.44^{+0.17}_{-0.15} \times 10^{-3} \evsqc$
and mixing angle,  $\sin^2(\theta_{23}) =0.514\pm0.082$.
We anticipate future T2K data will improve our neutrino disappearance
measurements, and our own measurements combined with other accelerator
and reactor measurements will lead to important constraints 
and more precise determinations of the fundamental neutrino 
mixing parameters.

We thank the J-PARC for the superb accelerator performance and the CERN NA61 collaboration for providing valuable particle production data.
We acknowledge the support of MEXT, Japan; 
NSERC, NRC and CFI, Canada;
CEA and CNRS/IN2P3, France;
DFG, Germany; 
INFN, Italy;
Ministry of Science and Higher Education, Poland;
RAS, RFBR and MES, Russia; 
MEST and NRF, South Korea;
MICINN and CPAN, Spain;
SNSF and SER, Switzerland;
STFC, U.K.; and 
DOE, U.S.A.
We also thank CERN for the UA1/NOMAD magnet, 
DESY for the HERA-B magnet mover system, and NII for SINET4.
In addition participation of individual researchers
and institutions has been further supported by funds from: ERC (FP7), EU; 
JSPS, Japan; 
Royal Society, UK; 
DOE Early Career program, U.S.A.


\bibliographystyle{unsrt}
\bibliographystyle{apsrev4-1}

\bibliography{arxiv-bibtex}

\providecommand{\noopsort}[1]{}\providecommand{\singleletter}[1]{#1}%
\begin{thebibliography}{40}%
\makeatletter
\providecommand \@ifxundefined [1]{%
 \@ifx{#1\undefined}
}%
\providecommand \@ifnum [1]{%
 \ifnum #1\expandafter \@firstoftwo
 \else \expandafter \@secondoftwo
 \fi
}%
\providecommand \@ifx [1]{%
 \ifx #1\expandafter \@firstoftwo
 \else \expandafter \@secondoftwo
 \fi
}%
\providecommand \natexlab [1]{#1}%
\providecommand \enquote  [1]{``#1''}%
\providecommand \bibnamefont  [1]{#1}%
\providecommand \bibfnamefont [1]{#1}%
\providecommand \citenamefont [1]{#1}%
\providecommand \href@noop [0]{\@secondoftwo}%
\providecommand \href [0]{\begingroup \@sanitize@url \@href}%
\providecommand \@href[1]{\@@startlink{#1}\@@href}%
\providecommand \@@href[1]{\endgroup#1\@@endlink}%
\providecommand \@sanitize@url [0]{\catcode `\\12\catcode `\$12\catcode
  `\&12\catcode `\#12\catcode `\^12\catcode `\_12\catcode `\%12\relax}%
\providecommand \@@startlink[1]{}%
\providecommand \@@endlink[0]{}%
\providecommand \url  [0]{\begingroup\@sanitize@url \@url }%
\providecommand \@url [1]{\endgroup\@href {#1}{\urlprefix }}%
\providecommand \urlprefix  [0]{URL }%
\providecommand \Eprint [0]{\href }%
\providecommand \doibase [0]{http://dx.doi.org/}%
\providecommand \selectlanguage [0]{\@gobble}%
\providecommand \bibinfo  [0]{\@secondoftwo}%
\providecommand \bibfield  [0]{\@secondoftwo}%
\providecommand \translation [1]{[#1]}%
\providecommand \BibitemOpen [0]{}%
\providecommand \bibitemStop [0]{}%
\providecommand \bibitemNoStop [0]{.\EOS\space}%
\providecommand \EOS [0]{\spacefactor3000\relax}%
\providecommand \BibitemShut  [1]{\csname bibitem#1\endcsname}%
\let\auto@bib@innerbib\@empty
\bibitem [{\citenamefont {Pontecorvo}(1957)}]{Pont1}%
  \BibitemOpen
  \bibfield  {author} {\bibinfo {author} {\bibfnamefont {B.}~\bibnamefont
  {Pontecorvo}},\ }\href@noop {} {\bibfield  {journal} {\bibinfo  {journal}
  {JETP}\ }\textbf {\bibinfo {volume} {6}},\ \bibinfo {pages} {429} (\bibinfo
  {year} {1957})}\BibitemShut {NoStop}%
\bibitem [{\citenamefont {Pontecorvo}(1958)}]{Pont2}%
  \BibitemOpen
  \bibfield  {author} {\bibinfo {author} {\bibfnamefont {B.}~\bibnamefont
  {Pontecorvo}},\ }\href@noop {} {\bibfield  {journal} {\bibinfo  {journal}
  {JETP}\ }\textbf {\bibinfo {volume} {7}},\ \bibinfo {pages} {172} (\bibinfo
  {year} {1958})}\BibitemShut {NoStop}%
\bibitem [{\citenamefont {Pontecorvo}(1968)}]{Pont3}%
  \BibitemOpen
  \bibfield  {author} {\bibinfo {author} {\bibfnamefont {B.}~\bibnamefont
  {Pontecorvo}},\ }\href@noop {} {\bibfield  {journal} {\bibinfo  {journal}
  {JETP}\ }\textbf {\bibinfo {volume} {26}},\ \bibinfo {pages} {984} (\bibinfo
  {year} {1968})}\BibitemShut {NoStop}%
\bibitem [{\citenamefont {Maki}\ \emph {et~al.}(1962)\citenamefont {Maki},
  \citenamefont {Nakagawa},\ and\ \citenamefont {Sakata}}]{MNS}%
  \BibitemOpen
  \bibfield  {author} {\bibinfo {author} {\bibfnamefont {Z.}~\bibnamefont
  {Maki}}, \bibinfo {author} {\bibfnamefont {M.}~\bibnamefont {Nakagawa}}, \
  and\ \bibinfo {author} {\bibfnamefont {S.}~\bibnamefont {Sakata}},\
  }\href@noop {} {\bibfield  {journal} {\bibinfo  {journal} {Prog. Theor.
  Phys.}\ }\textbf {\bibinfo {volume} {28}},\ \bibinfo {pages} {870} (\bibinfo
  {year} {1962})}\BibitemShut {NoStop}%
\bibitem [{\citenamefont {Beringer}\ and\ \citenamefont {others
  (PDG)}(2012)}]{MCMC_ref}%
  \BibitemOpen
  \bibfield  {author} {\bibinfo {author} {\bibfnamefont {J.}~\bibnamefont
  {Beringer}}\ and\ \bibinfo {author} {\bibnamefont {others (PDG)}},\
  }\href@noop {} {\bibfield  {journal} {\bibinfo  {journal} {Phys.Rev.}\
  }\textbf {\bibinfo {volume} {D86}},\ \bibinfo {pages} {010001} (\bibinfo
  {year} {2012})},\ \bibinfo {note} {we use the standard PDG notation for the
  neutrino mixing angles and see Section 37.5 for an introduction to Markov
  Chains and further references.},\ \Eprint
  {http://arxiv.org/abs/http://pdg.lbl.gov} {http://pdg.lbl.gov} \BibitemShut
  {NoStop}%
\bibitem [{\citenamefont {Abe}\ \emph {et~al.}(2012{\natexlab{a}})\citenamefont
  {Abe} \emph {et~al.}}]{PhysRevD.85.031103}%
  \BibitemOpen
  \bibfield  {author} {\bibinfo {author} {\bibfnamefont {K.}~\bibnamefont
  {Abe}} \emph {et~al.} (\bibinfo {collaboration} {The T2K Collaboration}),\
  }\href {\doibase 10.1103/PhysRevD.85.031103} {\bibfield  {journal} {\bibinfo
  {journal} {Phys. Rev. D}\ }\textbf {\bibinfo {volume} {85}},\ \bibinfo
  {pages} {031103} (\bibinfo {year} {2012}{\natexlab{a}})}\BibitemShut
  {NoStop}%
\bibitem [{\citenamefont {Wendell}\ \emph {et~al.}(2010)\citenamefont {Wendell}
  \emph {et~al.}}]{PhysRevD.81.092004}%
  \BibitemOpen
  \bibfield  {author} {\bibinfo {author} {\bibfnamefont {R.}~\bibnamefont
  {Wendell}} \emph {et~al.} (\bibinfo {collaboration} {The Super-Kamiokande
  Collaboration}),\ }\href {\doibase 10.1103/PhysRevD.81.092004} {\bibfield
  {journal} {\bibinfo  {journal} {Phys. Rev. D}\ }\textbf {\bibinfo {volume}
  {81}},\ \bibinfo {pages} {092004} (\bibinfo {year} {2010})}\BibitemShut
  {NoStop}%
\bibitem [{\citenamefont {Adamson}\ \emph {et~al.}(2012)\citenamefont {Adamson}
  \emph {et~al.}}]{Adamson:2012rm}%
  \BibitemOpen
  \bibfield  {author} {\bibinfo {author} {\bibfnamefont {P.}~\bibnamefont
  {Adamson}} \emph {et~al.} (\bibinfo {collaboration} {MINOS Collaboration}),\
  }\href {\doibase 10.1103/PhysRevLett.108.191801} {\bibfield  {journal}
  {\bibinfo  {journal} {Phys. Rev. Lett.}\ }\textbf {\bibinfo {volume} {108}},\
  \bibinfo {pages} {191801} (\bibinfo {year} {2012})}\BibitemShut {NoStop}%
\bibitem [{\citenamefont {Adamson}\ \emph {et~al.}(2013)\citenamefont {Adamson}
  \emph {et~al.}}]{minos-apr2013}%
  \BibitemOpen
  \bibfield  {author} {\bibinfo {author} {\bibfnamefont {P.}~\bibnamefont
  {Adamson}} \emph {et~al.} (\bibinfo {collaboration} {MINOS Collaboration}),\
  }\href@noop {} {\  (\bibinfo {year} {2013})},\ \Eprint
  {http://arxiv.org/abs/1304.6335} {arXiv:1304.6335 [hep-ex]} \BibitemShut
  {NoStop}%
\bibitem [{\citenamefont {Abe}\ \emph {et~al.}(2011{\natexlab{a}})\citenamefont
  {Abe} \emph {et~al.}}]{PhysRevLett.107.041801}%
  \BibitemOpen
  \bibfield  {author} {\bibinfo {author} {\bibfnamefont {K.}~\bibnamefont
  {Abe}} \emph {et~al.} (\bibinfo {collaboration} {T2K Collaboration}),\ }\href
  {\doibase 10.1103/PhysRevLett.107.041801} {\bibfield  {journal} {\bibinfo
  {journal} {Phys. Rev. Lett.}\ }\textbf {\bibinfo {volume} {107}},\ \bibinfo
  {pages} {041801} (\bibinfo {year} {2011}{\natexlab{a}})}\BibitemShut
  {NoStop}%
\bibitem [{\citenamefont {Adamson}\ \emph {et~al.}(2011)\citenamefont {Adamson}
  \emph {et~al.}}]{Adamson:2011qu}%
  \BibitemOpen
  \bibfield  {author} {\bibinfo {author} {\bibfnamefont {P.}~\bibnamefont
  {Adamson}} \emph {et~al.} (\bibinfo {collaboration} {MINOS Collaboration}),\
  }\href {\doibase 10.1103/PhysRevLett.107.181802} {\bibfield  {journal}
  {\bibinfo  {journal} {Phys. Rev. Lett.}\ }\textbf {\bibinfo {volume} {107}},\
  \bibinfo {pages} {181802} (\bibinfo {year} {2011})}\BibitemShut {NoStop}%
\bibitem [{\citenamefont {An}\ \emph {et~al.}(2012)\citenamefont {An} \emph
  {et~al.}}]{PhysRevLett.108.171803}%
  \BibitemOpen
  \bibfield  {author} {\bibinfo {author} {\bibfnamefont {F.~P.}\ \bibnamefont
  {An}} \emph {et~al.} (\bibinfo {collaboration} {Daya Bay Collaboration}),\
  }\href {\doibase 10.1103/PhysRevLett.108.171803} {\bibfield  {journal}
  {\bibinfo  {journal} {Phys. Rev. Lett.}\ }\textbf {\bibinfo {volume} {108}},\
  \bibinfo {pages} {171803} (\bibinfo {year} {2012})}\BibitemShut {NoStop}%
\bibitem [{\citenamefont {Abe}\ \emph {et~al.}(2012{\natexlab{b}})\citenamefont
  {Abe} \emph {et~al.}}]{PhysRevLett.108.131801}%
  \BibitemOpen
  \bibfield  {author} {\bibinfo {author} {\bibfnamefont {Y.}~\bibnamefont
  {Abe}} \emph {et~al.} (\bibinfo {collaboration} {Double Chooz
  Collaboration}),\ }\href {\doibase 10.1103/PhysRevLett.108.131801} {\bibfield
   {journal} {\bibinfo  {journal} {Phys. Rev. Lett.}\ }\textbf {\bibinfo
  {volume} {108}},\ \bibinfo {pages} {131801} (\bibinfo {year}
  {2012}{\natexlab{b}})}\BibitemShut {NoStop}%
\bibitem [{\citenamefont {Ahn}\ \emph {et~al.}(2012)\citenamefont {Ahn} \emph
  {et~al.}}]{PhysRevLett.108.191802}%
  \BibitemOpen
  \bibfield  {author} {\bibinfo {author} {\bibfnamefont {J.~K.}\ \bibnamefont
  {Ahn}} \emph {et~al.} (\bibinfo {collaboration} {RENO Collaboration}),\
  }\href {\doibase 10.1103/PhysRevLett.108.191802} {\bibfield  {journal}
  {\bibinfo  {journal} {Phys. Rev. Lett.}\ }\textbf {\bibinfo {volume} {108}},\
  \bibinfo {pages} {191802} (\bibinfo {year} {2012})}\BibitemShut {NoStop}%
\bibitem [{\citenamefont {Abe}\ \emph {et~al.}(2011{\natexlab{b}})\citenamefont
  {Abe} \emph {et~al.}}]{Abe:2011ks}%
  \BibitemOpen
  \bibfield  {author} {\bibinfo {author} {\bibfnamefont {K.}~\bibnamefont
  {Abe}} \emph {et~al.} (\bibinfo {collaboration} {T2K Collaboration}),\ }\href
  {\doibase 10.1016/j.nima.2011.06.067} {\bibfield  {journal} {\bibinfo
  {journal} {Nucl. Instrum. Methods Phys. Res., Sect. A}\ }\textbf {\bibinfo
  {volume} {659}},\ \bibinfo {pages} {106} (\bibinfo {year}
  {2011}{\natexlab{b}})},\ \bibinfo {note} {see Figure 16 for a schematic
  diagram of the ND280 detector.}\BibitemShut {Stop}%
\bibitem [{\citenamefont {Abe}\ \emph {et~al.}(2013{\natexlab{a}})\citenamefont
  {Abe} \emph {et~al.}}]{PhysRevD.87.012001}%
  \BibitemOpen
  \bibfield  {author} {\bibinfo {author} {\bibfnamefont {K.}~\bibnamefont
  {Abe}} \emph {et~al.} (\bibinfo {collaboration} {T2K Collaboration}),\ }\href
  {\doibase 10.1103/PhysRevD.87.012001} {\bibfield  {journal} {\bibinfo
  {journal} {Phys. Rev. D}\ }\textbf {\bibinfo {volume} {87}},\ \bibinfo
  {pages} {012001} (\bibinfo {year} {2013}{\natexlab{a}})},\ \bibinfo {note}
  {see the predicted flux at SK reweighted with the NA61/SHINE measurement in
  Fig. 39 and the neutrino events per POT as measured by the INGRID
  sub-detector in Fig. 12.}\BibitemShut {Stop}%
\bibitem [{\citenamefont {Abe}\ \emph {et~al.}(2012{\natexlab{c}})\citenamefont
  {Abe} \emph {et~al.}}]{Abe2012}%
  \BibitemOpen
  \bibfield  {author} {\bibinfo {author} {\bibfnamefont {K.}~\bibnamefont
  {Abe}} \emph {et~al.} (\bibinfo {collaboration} {T2K ND280 INGRID
  Collaboration}),\ }\href {\doibase 10.1016/j.nima.2012.03.023} {\bibfield
  {journal} {\bibinfo  {journal} {Nucl. Instrum. Methods Phys. Res., Sect. A}\
  }\textbf {\bibinfo {volume} {694}},\ \bibinfo {pages} {211} (\bibinfo {year}
  {2012}{\natexlab{c}})}\BibitemShut {NoStop}%
\bibitem [{\citenamefont {Assylbekov}\ \emph {et~al.}(2012)\citenamefont
  {Assylbekov} \emph {et~al.}}]{Assylbekov201248}%
  \BibitemOpen
  \bibfield  {author} {\bibinfo {author} {\bibfnamefont {S.}~\bibnamefont
  {Assylbekov}} \emph {et~al.} (\bibinfo {collaboration} {T2K ND280 P0D
  Collaboration}),\ }\href {\doibase 10.1016/j.nima.2012.05.028} {\bibfield
  {journal} {\bibinfo  {journal} {Nucl.Instrum.Meth.}\ }\textbf {\bibinfo
  {volume} {A686}},\ \bibinfo {pages} {48 } (\bibinfo {year}
  {2012})}\BibitemShut {NoStop}%
\bibitem [{\citenamefont {Abgrall}\ \emph
  {et~al.}(2011{\natexlab{a}})\citenamefont {Abgrall} \emph
  {et~al.}}]{Abgrall:2010hi}%
  \BibitemOpen
  \bibfield  {author} {\bibinfo {author} {\bibfnamefont {N.}~\bibnamefont
  {Abgrall}} \emph {et~al.} (\bibinfo {collaboration} {T2K ND280 TPC
  Collaboration}),\ }\href {\doibase 10.1016/j.nima.2011.02.036} {\bibfield
  {journal} {\bibinfo  {journal} {Nucl. Instrum. Methods Phys. Res., Sect. A}\
  }\textbf {\bibinfo {volume} {637}},\ \bibinfo {pages} {25} (\bibinfo {year}
  {2011}{\natexlab{a}})}\BibitemShut {NoStop}%
\bibitem [{\citenamefont {Amaudruz}\ \emph {et~al.}(2012)\citenamefont
  {Amaudruz} \emph {et~al.}}]{Amaudruz:2012pe}%
  \BibitemOpen
  \bibfield  {author} {\bibinfo {author} {\bibfnamefont {P.}~\bibnamefont
  {Amaudruz}} \emph {et~al.} (\bibinfo {collaboration} {T2K ND280 FGD
  Collaboration}),\ }\href {\doibase 10.1016/j.nima.2012.08.020} {\bibfield
  {journal} {\bibinfo  {journal} {Nucl. Instrum. Methods Phys. Res., Sect. A}\
  }\textbf {\bibinfo {volume} {696}},\ \bibinfo {pages} {1} (\bibinfo {year}
  {2012})}\BibitemShut {NoStop}%
\bibitem [{\citenamefont {Allan}\ \emph {et~al.}(2013)\citenamefont {Allan}
  \emph {et~al.}}]{allan2013electromagnetic}%
  \BibitemOpen
  \bibfield  {author} {\bibinfo {author} {\bibfnamefont {D.}~\bibnamefont
  {Allan}} \emph {et~al.},\ }\href@noop {} {\bibfield  {journal} {\bibinfo
  {journal} {arXiv preprint arXiv:1308.3445}\ } (\bibinfo {year}
  {2013})}\BibitemShut {NoStop}%
\bibitem [{\citenamefont {Aoki}\ \emph {et~al.}(2013)\citenamefont {Aoki} \emph
  {et~al.}}]{Aoki:2012mf}%
  \BibitemOpen
  \bibfield  {author} {\bibinfo {author} {\bibfnamefont {S.}~\bibnamefont
  {Aoki}} \emph {et~al.} (\bibinfo {collaboration} {T2K ND280 SMRD
  Collaboration}),\ }\href {\doibase 10.1016/j.nima.2012.10.001} {\bibfield
  {journal} {\bibinfo  {journal} {Nucl.Instrum.Meth.}\ }\textbf {\bibinfo
  {volume} {A698}},\ \bibinfo {pages} {135} (\bibinfo {year} {2013})},\ \Eprint
  {http://arxiv.org/abs/1206.3553} {arXiv:1206.3553 [physics.ins-det]}
  \BibitemShut {NoStop}%
\bibitem [{\citenamefont {Ashie}\ \emph {et~al.}(2005)\citenamefont {Ashie}
  \emph {et~al.}}]{Ashie:2005ik}%
  \BibitemOpen
  \bibfield  {author} {\bibinfo {author} {\bibfnamefont {Y.}~\bibnamefont
  {Ashie}} \emph {et~al.} (\bibinfo {collaboration} {Super-Kamiokande
  Collaboration}),\ }\href {\doibase 10.1103/PhysRevD.71.112005} {\bibfield
  {journal} {\bibinfo  {journal} {Phys. Rev. D}\ }\textbf {\bibinfo {volume}
  {71}},\ \bibinfo {pages} {112005} (\bibinfo {year} {2005})}\BibitemShut
  {NoStop}%
\bibitem [{\citenamefont {Abgrall}\ \emph
  {et~al.}(2011{\natexlab{b}})\citenamefont {Abgrall} \emph
  {et~al.}}]{Abgrall:2011ae}%
  \BibitemOpen
  \bibfield  {author} {\bibinfo {author} {\bibfnamefont {N.}~\bibnamefont
  {Abgrall}} \emph {et~al.} (\bibinfo {collaboration} {NA61/SHINE
  Collaboration}),\ }\href {\doibase 10.1103/PhysRevC.84.034604} {\bibfield
  {journal} {\bibinfo  {journal} {Phys. Rev. C}\ }\textbf {\bibinfo {volume}
  {84}},\ \bibinfo {pages} {034604} (\bibinfo {year}
  {2011}{\natexlab{b}})}\BibitemShut {NoStop}%
\bibitem [{\citenamefont {Abgrall}\ \emph {et~al.}(2012)\citenamefont {Abgrall}
  \emph {et~al.}}]{Abgrall:2011ts}%
  \BibitemOpen
  \bibfield  {author} {\bibinfo {author} {\bibfnamefont {N.}~\bibnamefont
  {Abgrall}} \emph {et~al.} (\bibinfo {collaboration} {NA61/SHINE
  Collaboration}),\ }\href {\doibase 10.1103/PhysRevC.85.035210} {\bibfield
  {journal} {\bibinfo  {journal} {Phys. Rev. C}\ }\textbf {\bibinfo {volume}
  {85}},\ \bibinfo {pages} {035210} (\bibinfo {year} {2012})}\BibitemShut
  {NoStop}%
\bibitem [{\citenamefont {Ferrari}\ \emph {et~al.}()\citenamefont {Ferrari},
  \citenamefont {Sala}, \citenamefont {Fasso},\ and\ \citenamefont
  {Ranft}}]{Ferrari:2005zk}%
  \BibitemOpen
  \bibfield  {author} {\bibinfo {author} {\bibfnamefont {A.}~\bibnamefont
  {Ferrari}}, \bibinfo {author} {\bibfnamefont {P.~R.}\ \bibnamefont {Sala}},
  \bibinfo {author} {\bibfnamefont {A.}~\bibnamefont {Fasso}}, \ and\ \bibinfo
  {author} {\bibfnamefont {J.}~\bibnamefont {Ranft}},\ }\href@noop {} {\bibinfo
   {journal} {CERN-2005-010, SLAC-R-773, INFN-TC-05-11}\ }\BibitemShut
  {NoStop}%
\bibitem [{\citenamefont {Battistoni}\ \emph {et~al.}(2007)\citenamefont
  {Battistoni}, \citenamefont {Muraro}, \citenamefont {Sala}, \citenamefont
  {Cerutti}, \citenamefont {Ferrari} \emph {et~al.}}]{Battistoni:2007zzb}%
  \BibitemOpen
\bibfield  {journal} {  }\bibfield  {author} {\bibinfo {author} {\bibfnamefont
  {G.}~\bibnamefont {Battistoni}}, \bibinfo {author} {\bibfnamefont
  {S.}~\bibnamefont {Muraro}}, \bibinfo {author} {\bibfnamefont {P.~R.}\
  \bibnamefont {Sala}}, \bibinfo {author} {\bibfnamefont {F.}~\bibnamefont
  {Cerutti}}, \bibinfo {author} {\bibfnamefont {A.}~\bibnamefont {Ferrari}},
  \emph {et~al.},\ }\href {\doibase 10.1063/1.2720455} {\bibfield  {journal}
  {\bibinfo  {journal} {AIP Conf.Proc.}\ }\textbf {\bibinfo {volume} {896}},\
  \bibinfo {pages} {31} (\bibinfo {year} {2007})},\ \bibinfo {note} {we used
  FLUKA2008, which was the latest version at the time of this
  study.}\BibitemShut {Stop}%
\bibitem [{\citenamefont {Brun}\ \emph {et~al.}(1994)\citenamefont {Brun},
  \citenamefont {Carminati},\ and\ \citenamefont {Giani}}]{GEANT3}%
  \BibitemOpen
  \bibfield  {author} {\bibinfo {author} {\bibfnamefont {R.}~\bibnamefont
  {Brun}}, \bibinfo {author} {\bibfnamefont {F.}~\bibnamefont {Carminati}}, \
  and\ \bibinfo {author} {\bibfnamefont {S.}~\bibnamefont {Giani}},\
  }\href@noop {} {\bibfield  {journal} {\bibinfo  {journal} {CERN-W5013}\ }
  (\bibinfo {year} {1994})}\BibitemShut {NoStop}%
\bibitem [{\citenamefont {Zeitnitz}\ and\ \citenamefont
  {Gabriel}(1993)}]{GCALOR}%
  \BibitemOpen
  \bibfield  {author} {\bibinfo {author} {\bibfnamefont {C.}~\bibnamefont
  {Zeitnitz}}\ and\ \bibinfo {author} {\bibfnamefont {T.~A.}\ \bibnamefont
  {Gabriel}},\ }\href@noop {} {\bibfield  {journal} {\bibinfo  {journal} {In
  Proc. of International Conference on Calorimetry in High Energy Physics}\ }
  (\bibinfo {year} {1993})}\BibitemShut {NoStop}%
\bibitem [{\citenamefont {Abe}\ \emph {et~al.}(2013{\natexlab{b}})\citenamefont
  {Abe} \emph {et~al.}}]{nue2013}%
  \BibitemOpen
  \bibfield  {author} {\bibinfo {author} {\bibfnamefont {K.}~\bibnamefont
  {Abe}} \emph {et~al.} (\bibinfo {collaboration} {T2K Collaboration}),\
  }\href@noop {} {\bibfield  {journal} {\bibinfo  {journal} {submitted to Phys.
  Rev. D, see the sections IV. and V. on the neutrino interaction model and the
  neutrino flux model}\ } (\bibinfo {year} {2013}{\natexlab{b}})},\ \Eprint
  {http://arxiv.org/abs/1304.0841} {arXiv:1304.0841 [hep-ph]} \BibitemShut
  {NoStop}%
\bibitem [{\citenamefont {Hayato}(2009)}]{Hayato:2009}%
  \BibitemOpen
  \bibfield  {author} {\bibinfo {author} {\bibfnamefont {Y.}~\bibnamefont
  {Hayato}},\ }\href@noop {} {\bibfield  {journal} {\bibinfo  {journal} {Acta
  Phys. Pol. B}\ }\textbf {\bibinfo {volume} {40}},\ \bibinfo {pages} {2477}
  (\bibinfo {year} {2009})}\BibitemShut {NoStop}%
\bibitem [{\citenamefont {Aguilar-Arevalo}\ \emph {et~al.}(2010)\citenamefont
  {Aguilar-Arevalo} \emph {et~al.}}]{mb-ccqe}%
  \BibitemOpen
  \bibfield  {author} {\bibinfo {author} {\bibfnamefont {A.~A.}\ \bibnamefont
  {Aguilar-Arevalo}} \emph {et~al.} (\bibinfo {collaboration} {MiniBooNE
  Collaboration}),\ }\href {\doibase 10.1103/PhysRevD.81.092005} {\bibfield
  {journal} {\bibinfo  {journal} {Phys. Rev. D}\ }\textbf {\bibinfo {volume}
  {81}},\ \bibinfo {pages} {092005} (\bibinfo {year} {2010})}\BibitemShut
  {NoStop}%
\bibitem [{\citenamefont {Nieves}\ \emph {et~al.}(2012)\citenamefont {Nieves},
  \citenamefont {Simo},\ and\ \citenamefont {Vacas}}]{Nieves}%
  \BibitemOpen
  \bibfield  {author} {\bibinfo {author} {\bibfnamefont {J.}~\bibnamefont
  {Nieves}}, \bibinfo {author} {\bibfnamefont {I.~R.}\ \bibnamefont {Simo}}, \
  and\ \bibinfo {author} {\bibfnamefont {M.~V.}\ \bibnamefont {Vacas}},\ }\href
  {\doibase 10.1016/j.physletb.2011.11.061} {\bibfield  {journal} {\bibinfo
  {journal} {Phys.Lett.B}\ }\textbf {\bibinfo {volume} {707}},\ \bibinfo
  {pages} {72} (\bibinfo {year} {2012})}\BibitemShut {NoStop}%
\bibitem [{\citenamefont {Martini}\ \emph {et~al.}(2010)\citenamefont
  {Martini}, \citenamefont {Ericson}, \citenamefont {Chanfray},\ and\
  \citenamefont {Marteau}}]{Martini:2010}%
  \BibitemOpen
  \bibfield  {author} {\bibinfo {author} {\bibfnamefont {M.}~\bibnamefont
  {Martini}}, \bibinfo {author} {\bibfnamefont {M.}~\bibnamefont {Ericson}},
  \bibinfo {author} {\bibfnamefont {G.}~\bibnamefont {Chanfray}}, \ and\
  \bibinfo {author} {\bibfnamefont {J.}~\bibnamefont {Marteau}},\ }\href
  {\doibase 10.1103/PhysRevC.81.045502} {\bibfield  {journal} {\bibinfo
  {journal} {Phys. Rev. C}\ }\textbf {\bibinfo {volume} {81}},\ \bibinfo
  {pages} {045502} (\bibinfo {year} {2010})}\BibitemShut {NoStop}%
\bibitem [{\citenamefont {Meloni}\ and\ \citenamefont
  {Martini}(2012)}]{Meloni}%
  \BibitemOpen
  \bibfield  {author} {\bibinfo {author} {\bibfnamefont {D.}~\bibnamefont
  {Meloni}}\ and\ \bibinfo {author} {\bibfnamefont {M.}~\bibnamefont
  {Martini}},\ }\href {\doibase 10.1016/j.physletb.2012.08.007} {\bibfield
  {journal} {\bibinfo  {journal} {Phys.Lett.B}\ }\textbf {\bibinfo {volume}
  {716}},\ \bibinfo {pages} {186} (\bibinfo {year} {2012})},\ \Eprint
  {http://arxiv.org/abs/1203.3335} {arXiv:1203.3335 [hep-ex]} \BibitemShut
  {NoStop}%
\bibitem [{\citenamefont {Lalakulich}\ \emph {et~al.}(2012)\citenamefont
  {Lalakulich}, \citenamefont {Mosel},\ and\ \citenamefont
  {Gallmeister}}]{Lalakulich3}%
  \BibitemOpen
  \bibfield  {author} {\bibinfo {author} {\bibfnamefont {O.}~\bibnamefont
  {Lalakulich}}, \bibinfo {author} {\bibfnamefont {U.}~\bibnamefont {Mosel}}, \
  and\ \bibinfo {author} {\bibfnamefont {K.}~\bibnamefont {Gallmeister}},\
  }\href {\doibase 10.1103/PhysRevC.86.054606} {\bibfield  {journal} {\bibinfo
  {journal} {Phys.Rev. C}\ }\textbf {\bibinfo {volume} {86}},\ \bibinfo {pages}
  {054606} (\bibinfo {year} {2012})},\ \Eprint {http://arxiv.org/abs/1208.3678}
  {arXiv:1208.3678 [nucl-th]} \BibitemShut {NoStop}%
\bibitem [{\citenamefont {Martini}\ \emph {et~al.}(2012)\citenamefont
  {Martini}, \citenamefont {Ericson},\ and\ \citenamefont
  {Chanfray}}]{Martini2}%
  \BibitemOpen
  \bibfield  {author} {\bibinfo {author} {\bibfnamefont {M.}~\bibnamefont
  {Martini}}, \bibinfo {author} {\bibfnamefont {M.}~\bibnamefont {Ericson}}, \
  and\ \bibinfo {author} {\bibfnamefont {G.}~\bibnamefont {Chanfray}},\ }\href
  {\doibase 10.1103/PhysRevD.85.093012} {\bibfield  {journal} {\bibinfo
  {journal} {Phys.Rev.D}\ }\textbf {\bibinfo {volume} {85}},\ \bibinfo {pages}
  {093012} (\bibinfo {year} {2012})},\ \Eprint {http://arxiv.org/abs/1202.4745}
  {arXiv:1202.4745 [hep-ph]} \BibitemShut {NoStop}%
\bibitem [{\citenamefont {Martini}\ \emph {et~al.}(2013)\citenamefont
  {Martini}, \citenamefont {Ericson},\ and\ \citenamefont
  {Chanfray}}]{Martini3}%
  \BibitemOpen
  \bibfield  {author} {\bibinfo {author} {\bibfnamefont {M.}~\bibnamefont
  {Martini}}, \bibinfo {author} {\bibfnamefont {M.}~\bibnamefont {Ericson}}, \
  and\ \bibinfo {author} {\bibfnamefont {G.}~\bibnamefont {Chanfray}},\
  }\href@noop {} {\bibfield  {journal} {\bibinfo  {journal} {Phys.Rev.D}\
  }\textbf {\bibinfo {volume} {87}},\ \bibinfo {pages} {013009} (\bibinfo
  {year} {2013})},\ \Eprint {http://arxiv.org/abs/1211.1523v2}
  {arXiv:1211.1523v2 [hep-ph]} \BibitemShut {NoStop}%
\bibitem [{\citenamefont {Hagiwara}\ \emph {et~al.}(2011)\citenamefont
  {Hagiwara}, \citenamefont {Okamura},\ and\ \citenamefont
  {Senda}}]{Hagiwara:2011kw}%
  \BibitemOpen
  \bibfield  {author} {\bibinfo {author} {\bibfnamefont {K.}~\bibnamefont
  {Hagiwara}}, \bibinfo {author} {\bibfnamefont {N.}~\bibnamefont {Okamura}}, \
  and\ \bibinfo {author} {\bibfnamefont {K.-i.}\ \bibnamefont {Senda}},\ }\href
  {\doibase 10.1007/JHEP09(2011)082} {\bibfield  {journal} {\bibinfo  {journal}
  {JHEP}\ }\textbf {\bibinfo {volume} {1109}},\ \bibinfo {pages} {082}
  (\bibinfo {year} {2011})},\ \Eprint {http://arxiv.org/abs/1107.5857}
  {arXiv:1107.5857 [hep-ph]} \BibitemShut {NoStop}%
\bibitem [{\citenamefont {Itow}(2013)}]{Itow201379}%
  \BibitemOpen
  \bibfield  {author} {\bibinfo {author} {\bibfnamefont {Y.}~\bibnamefont
  {Itow}},\ }\href {\doibase 10.1016/j.nuclphysbps.2013.03.014} {\bibfield
  {journal} {\bibinfo  {journal} {Nuclear Physics B, Proceedings Supplements}\
  }\textbf {\bibinfo {volume} {235}},\ \bibinfo {pages} {79} (\bibinfo {year}
  {2013})},\ \bibinfo {note} {the \{XXV\} International Conference on Neutrino
  Physics and Astrophysics}\BibitemShut {NoStop}%
\end{thebibliography}%

\end{document}